\documentclass[twocolumn]{openjournal}

\usepackage{xcolor}
\usepackage[utf8]{inputenc}
\usepackage[english]{babel}
\usepackage{graphicx}
\usepackage{amsmath}
\usepackage{orcidlink}
\usepackage{multirow}
\usepackage{color,colortbl}
\DeclareGraphicsExtensions{.bmp,.png,.jpg,.pdf}
\usepackage{verbatim}
\definecolor{linkcolor}{rgb}{0.0,0.3,0.5}
\usepackage{hyperref}
\hypersetup{
    unicode, 
    colorlinks=true,
    linkcolor=linkcolor,
    citecolor=linkcolor,
    filecolor=linkcolor,
    urlcolor=linkcolor,
}
\newcommand{\reb}{\texttt{REBOUND}}
\newcommand{\magrathea}{\texttt{MAGRATHEA}}
\newcommand{\mercurius}{\texttt{MERCURIUS}}
\newcommand{\Nbody}{\textit{N}-body}
\newcommand{\grains}{\texttt{GRAINS}}

\begin{document}

\title{Correlations between the Carbon-to-Oxygen Ratio and the Orbital and Physical Properties of Exoplanets} 

\author{Noah Ferich$^{1,}$$^2$\orcidlink{0000-0001-8140-1727}}
\author{Robert Royer III$^{1,}$$^2$\orcidlink{0000-0002-0900-0192}}
\author{Cody Shakespeare$^{1,}$$^2$\orcidlink{0000-0003-4709-2689}}
\author{Jason H. Steffen$^{1,}$$^2$\orcidlink{0000-0003-2202-3847}}

\affiliation{$^1$Department of Physics and Astronomy, University of Nevada, Las Vegas, 4505 South Maryland Parkway, Las Vegas, NV 89154, USA}
\affiliation{$^2$Nevada Center for Astrophysics, University of Nevada, Las Vegas, 4505 South Maryland Parkway, Las Vegas, NV 89154, USA}

\email{fericn1@unlv.nevada.edu}

\begin{abstract}
 Studies suggest that protoplanetary disks around stars with high carbon-to-oxygen (C/O) ratios can produce terrestrial planets with carbon-dominated compositions instead of silicon-dominated ones. While previous work studied the evolution of these disks and the planets that form within them, none analyze in detail how the differences in the surface density profiles of these disks could affect the architectures and iron and siderophile contents of exoplanetary systems. Here, we present a new study that uses the N-body integrator REBOUND and data from a sequential dust condensation model to simulate the formation of planets in systems with varying C/O ratios. We find that bulk density could be used to distinguish between silicon-rich and carbon-rich planets, though it may not be a reliable metric on its own.
\end{abstract}

\section{Introduction}
\label{sec:introduction}

Previous studies indicate that rocky planets are likely common throughout the galaxy \citep{Bryson2021} and can be enriched or depleted in certain chemical species compared to the inner four planets of the Solar System \citep{Bond2010}. Researchers often link these compositional discrepancies to a host star's metallicity since this parameter likely reflects the conditions of the protoplanetary disk in which these exoplanets formed \citep{Adibekyan2021}. Of particular interest has been the C/O ratios of these stars and systems, which theories suggest can determine whether or not a planetary system has a primarily silicate composition --- like the inner Solar System --- or a composition dominated by carbon and its compounds.

While the vast majority of nearby stars appear to have C/O ratios below unity \citep{Brewer&Fischer2016, Stonkut2020, Delgado2021, Sharma2024}, stars with ratios above unity do exist,\footnote{\citet{Brewer&Fischer2016} found $<4$ in a sample of 10,000 stars.} making the discovery of carbon planets rare but possible. So far, only one candidate around a main-sequence star, 55 Cancri e, has been found \citep{Madhusudhan2012}. However, separate studies of its host star's C/O ratio have returned several different values \citep{Delgado_Mena2010, Teske2013, Brewer&Fischer2016}, some of which are substantially less than unity, leading to uncertainty about its carbon-rich nature.

Considering the systematic errors present in trying to measure C/O ratios in stellar atmospheres \citep{Teske2013, Sharma2024}, we require other methods to determine if a planet is carbon-rich. For example, if the carbon content of a planet is related to its core mass fraction (CMF) or bulk density, then we could differentiate between carbon and silicon planets using measurements of their masses and radii. In addition, protoplanetary disks with different C/O ratios tend to have different surface density profiles \citep{Shakespeare2025}, which could lead to planets forming at different distances around a star and with different system architectures. If the differences between these systems are significant enough, then we could use them to classify planets as silicate or carbon rich.

Previous work has developed multiple models and pieces of software to resolve the chemical compositions and structure of early protoplanetary disks \citep{Cassen1996, Ciesla&Fred2008, Bond2010, Bond&Lauretta&Obrien2010, Pignatale2016, Li2020}. In addition, numerous studies have used \Nbody\ simulations to simulate the formation of planetary systems, especially during the late-stage of planet formation where interactions are dominated by gravity and impacts between protoplanets and planetesimals \citep{Kokubo2010, Chambers2013, Bonsor2015, Childs2019, Scora2020, childs2022, Scora2022, Kaufman2023, Childs2023, Scora2024}. These simulations are often used to perform statistical studies on the architecture and multiplicity of planetary systems and the physical properties of the resulting planets, including their sizes, masses, and compositions. Researchers have added additional features to these simulations in recent years, including imperfect collisions between particles, which can affect the final masses and compositions of the resulting planets. Some studies have connected these chemical and \Nbody\ simulations \citep{Bond2010, Bond&Lauretta&Obrien2010, Bond&Obrien&Lauretta2010, Elser2012, Moriarty2014} to investigate the compositional diversity of the planets that form. Here, we will perform a similar investigation using state-of-the-art models and software with a focus on the C/O ratios of these systems.

In this paper, we use surface density profiles from \citet{Shakespeare2025} produced by the sequential dust condensation model in \citet{Li2020} to create late-stage protoplanetary disks with varying C/O ratios around a Sun-like star. We simulate the formation of planets from these disks using the \Nbody\ software \reb\ \citep{Rein2012} and include the fragmentation module for \reb\ created by \citet{childs2022} to implement imperfect collisions between forming planets in these simulations. In addition, we use the post-processing Differentiated Body Composition Tracker (DBCT) from \citep{Ferich2025} to track how the CMFs of objects change over the course of each run and use the interior modelling code \magrathea\ \citep{Li2020} to estimate their bulk densities. With these results, we try to discern any differences between carbon-rich and silicon-rich systems and planets and determine whether are not they are significant enough to be seen in observations. 

In Section \ref{sec:models_and_methods}, we describe how we used sequential condensation data to create each disk, how we implemented these disks into \reb\ simulations, and how we determined the CMFs of planets produced in each simulation. In Section \ref{sec:results}, we analyze the different planetary systems and determine whether or not the planets' physical and orbital properties are related to the initial disk's C/O ratio and analyze the CMFs and bulk densities of these planets using \magrathea\ to see if there are differences between carbon and silicate planets. Finally, in section \ref{sec:discussion}, we discuss whether or not our results produced observable differences and describe how our methods and assumptions can be improved. 

\section{Models and Methods}
\label{sec:models_and_methods}

\subsection{Sequential Condensation Model}
\label{sec:sequential_condensation_model}

We use results produced by the sequential dust condensation model found in \citet{Li2020}. This model combines the disk evolution model found in \citet{Cassen1996}, the dust advection model in \citet{Jacquet2012}, and the \grains\ condensation code \citet{Petaev2009} to allow chemical species to sequentially condense out of a disk rather than all at once. Their algorithm uses three different phases for its over 500 condensates: a gaseous phase, an advected dust phase that allows condensates to flow with gases in the disk, and a decoupled dust phase that allows condensates to settle and remain unperturbed at a single orbital distance. This algorithm allows for condensing species and the evolving disk to react with each other in a feedback loop to produce the desired sequential dust production. See \citet{Li2020} for more details.

The data for this paper come from \citet{Shakespeare2025}, who used this model to study how C/O ratios affect the compositions of protoplanetary disks around Sun-like stars. They used two methods to change the C/O ratio in their simulations: varying the amount of carbon and oxygen in a simulation while keeping their sum constant and varying the amount of carbon while keeping the amount of oxygen constant. The data we use were produced using the first method, which they defined as the ``constant sum'' method in their publication. Note that the amounts of other elements are held constant at solar values, so this method will change the ratios of other elements with carbon and oxygen such as the nitrogen to oxygen ratio. We assume that this will not affect our results significantly.

\citet{Shakespeare2025} showed that the transition between silicate-dominated and carbon-dominated system occurs at C/O ratios between 0.8-1.0. Therefore, we choose C/O ratios to study that lie below, within, and above this range: 0.5, 0.7, 0.8, 0.9, 1.0, 1.1, 1.5, and 2.0. For reference, the Sun's C/O ratio is around 0.52 and is considered average when it comes to the solar neighborhood \citep{Lodders2003, Brewer&Fischer2016}. Previous surveys have not found any stars with C/O ratios of 1.5 or 2.0, making our exploration of these systems academic in nature.

\begin{figure}
	\includegraphics[width=\columnwidth]{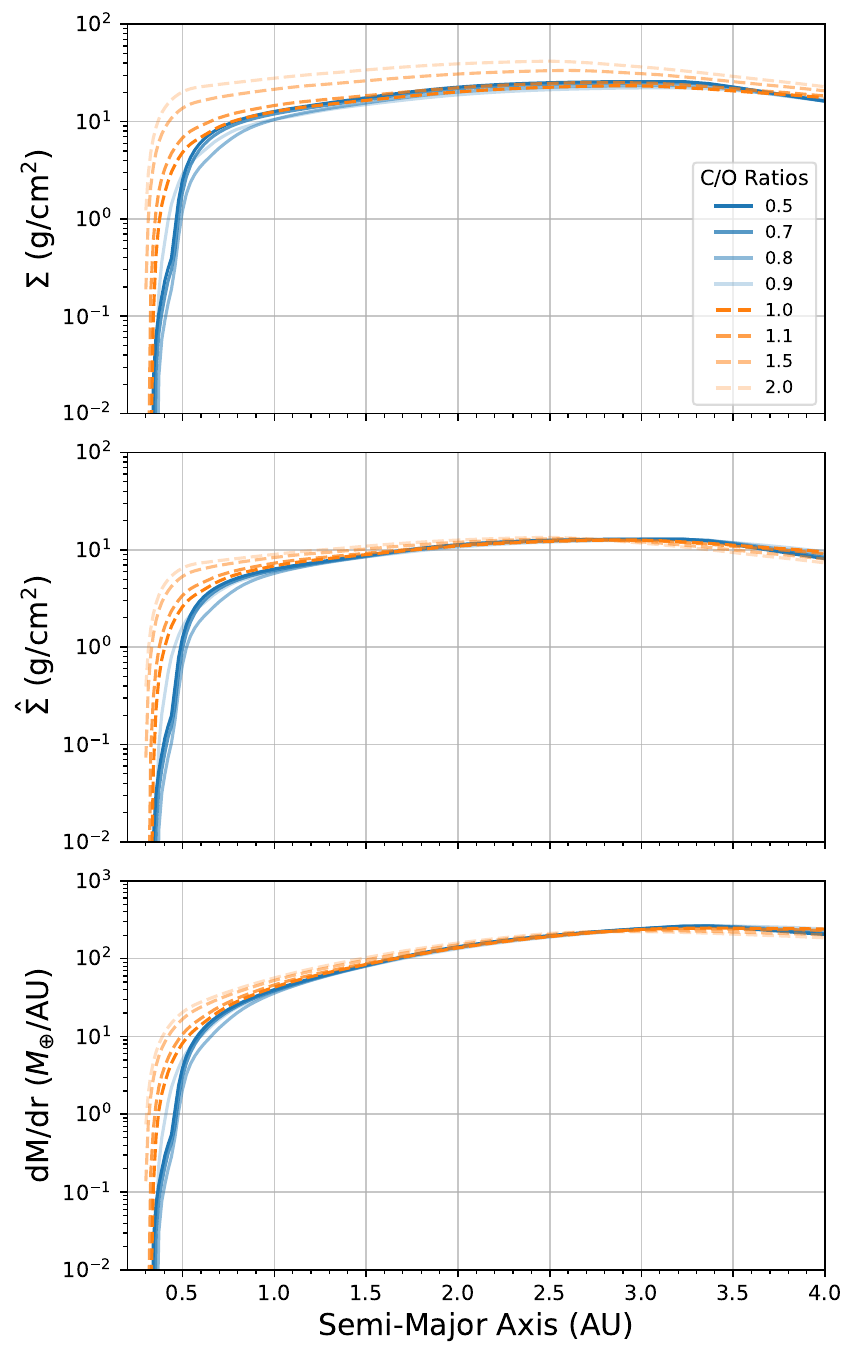}
    \caption{Surface densities and available mass within 0.3 AU to 4.0 AU in protoplanetary disks with different C/O ratios. The top panel shows the surface density at a particular distance, the middle panel shows the surface density profiles from the top panel when each disk is normalized to have a total of 10.0 $M_{\oplus}$, and the bottom panel shows the rate of change between total mass and distance for each disk assuming they are circular and thin. The color of a line indicates if the C/O ratio is low or high, and its opacity indicates the exact value.}
    \label{fig:surface_density_plot}
\end{figure}

Some of the results from \citet{Shakespeare2025} provide total surface density values for different regions in the disk. By interpolating these individual points, we create a surface density profile that can be used to determine the placement of planetary embryos and planetesimals in the disk (see Section \ref{sec:initial_disks}). The top panel of Figure \ref{fig:surface_density_plot} shows the total surface density profiles we produced. The profiles extend past 10 AU; however, we consider only the portion from 0.3 AU to 4 AU, which is the region where we place objects in our \Nbody\ simulations. Solid lines show C/O ratios below unity, which we refer to as low C/O ratios, and the dashed lines are C/O ratios above or at unity, which we refer to as high C/O ratios. For this study, we assume that low C/O ratios will produce silicon-dominated planets and high C/O ratios will produce carbon planets.

The plot shows that high C/O ratio disks have larger surface densities near their host stars than low C/O ratios. However, after a distance of around 1.0 AU, the density profiles for all ratios become similar in shape and value. The only two exceptions are with C/O ratios of 1.5 and 2.0, which have substantially larger densities at effectively all points in the disk. These results indicate that higher C/O ratio disks are likely to have more mass available for planet formation and will form more massive planets, a point we discuss in Section \ref{sec:discussion}. However, as the next section will describe, we normalize the surface density profiles so that all disks will have the same total mass.

\subsection{Initial Disks}
\label{sec:initial_disks}

When determining the placement of objects for our simulations, we assume that the protoplanetary disks from which they form are all circular and perfectly symmetrical about the system's host star and extend from 0.3 AU to 4.0 AU. Additionally, in order to form a variety of Super-Earths and nothing larger, we assume that every disk contains 20 $M_{\oplus}$ of material. The surface density profiles shown in the top panel of Figure \ref{fig:surface_density_plot} indicates that disks with different C/O ratios will not have the same total mass. Therefore, we normalize each of these profiles so that they contain the chosen amount of material. The center panel of Figure \ref{fig:surface_density_plot} shows the surface density profiles for each C/O ratio after normalization.

Following the procedure in \citet{Quintana2016}, \citet{childs2022}, and \citet{Ferich2025}, we create a late-stage protoplanetary disk with 260 planetesimals and 26 embryos that are ten times more massive than the planetesimals. For this study, our planetesimal mass is 0.0385 $M_{\oplus}$ and embryo mass is 0.385 $M_{\oplus}$. With this arrangement, the total mass is split evenly between planetesimals and embryos. To determine where objects are placed in the disk, we start at the inner edge of the disk and move outwards until a particular region has enough mass to form a planetesimal or embryo, which we determine by numerically integrating the curves shown in the bottom panel of Figure \ref{fig:surface_density_plot}. We place the embryo at the midpoint between the inner edge and outer edge of the region and begin this process again at the outer edge.

For example, to create the first embryo in a particular disk, we start at 0.3 AU and move outwards until we encompass a region that contains 0.385 $M_{\oplus}$. If we assume this region occurs between 0.3 AU to 0.7 AU, then we place the embryo at 0.5 AU and repeat this method, now starting at 0.7 AU and continuing outwards until the entire disk contains 26 planetary embryos. We then perform this process over again for planetesimals until the disk contains 260 of them.

Once the semi-major axes of all objects have been determined, we draw random values from uniform distributions for other orbital parameters. Eccentricities and inclinations can have the following ranges: $0 < e < 0.01$ and $0^\circ < i < 1^\circ$. In addition, the argument of periastron ($\omega$), the mean anomaly ($M$), and the longitude of ascending node ($\Omega$) are all chosen from a distribution that goes from $0^\circ$ to $360^\circ$. 

We perform all of these steps for each C/O ratio until we have 10 disks with identical semi-major axes but randomized properties for all other orbital parameters, giving us 80 disks total. In addition, we take the objects in these disks and scale their semi-major axes down by a factor of 5 so that they all lie between 0.09 AU and 0.8 AU, giving us another 80 disks. The original disks, which are optimized to produce more solar system analogues, we call ``regular disks,'' and the smaller disks, which are optimized to produce more Kepler-like systems with shorter orbital periods, we call ``compact'' disks. 

\begin{figure}
	\includegraphics[width=\columnwidth]{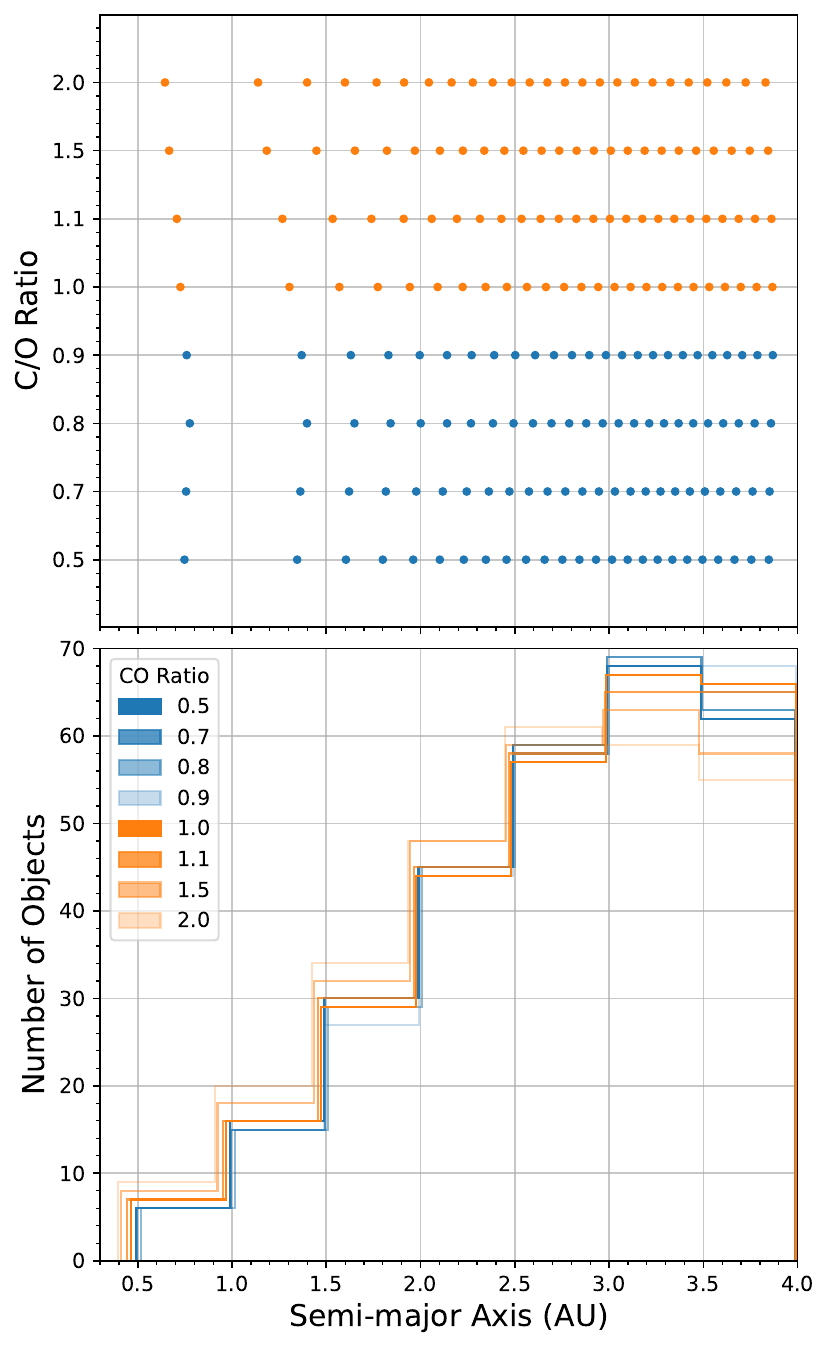}
    \caption{How objects are distributed in a regular disk depending on the C/O ratio. The top panel shows where the planetary embryos are placed in each disk. The bottom panel is a histogram showing how many objects (embryos and planetesimals) are contained in certain regions of each disk. We note that the compact disks will have the exact same distributions of material except scaled down into a region within 0.09 AU to 0.8 AU.}
    \label{fig:regular_disk_distribution}
\end{figure}

The top panel of Figure \ref{fig:regular_disk_distribution} shows the locations for all embryos in our regular disks. Similar to the normalized surface density profiles in Figure \ref{fig:surface_density_plot}, high C/O ratio disks seem to have more embryos at shorter distances and fewer at large distances when compared to low C/O ratio disks, though the effect is small. This minor discrepancy also appears in the bottom panel of the figure, which shows the cumulative distributions of all objects in each regular disk. In this case, the difference is most prominent for the highest two C/O ratios and less distinct for the ratios 1.1 and 1.0. We choose not show these distributions for compact disks because they are identical in shape to the ones displayed for regular disks --- the only difference being their semi-major axes are scaled down to a range between 0.09 AU and 0.8 AU.

\subsection{N-body Simulations and Fragmentation}
\label{sec:nbody_simulations_and_fragmentation}

We use \reb\ \citep{Rein2012} to simulate the formation of planetary systems around a 1 $M_{\odot}$ star from 160 initial disks. Each simulation is run for 10 Myr of simulation time or until they reach 1152 hours of CPU time. In certain cases, simulations did not reach the allotted 10 Myr of simulation time. However, in all cases, we were able to obtain full planetary systems with distinct planets and some leftover material. We use the \mercurius\ integrator, which can accurately treat close encounters between objects, for every simulation. For regular disks, we use a time step of 0.01 years, which corresponds to around 1/25 of the period of the innermost orbit, and for compact disks, we use a time step of 0.001 years, which corresponds to around 1/22 of the period of the innermost orbit. We also use an expansion factor of 3 on all particles in all simulations. This factor artificially inflates the initial radius of particles at the start of the integration. We note that this parameter can affect the collisional history of planets \citep{childs2022} and how much radial mixing occurs in the system during formation \citep{Grimm2026}; we assume our value for it is small enough to keep these effects to a minimum.

We utilize the fragmentation module for \reb\ created by \citet{childs2022} to implement imperfect collisions between the objects. This module uses models from \citet{Leinhardt2012}, \citet{Asphaug2010}, and \citet{Genda2012} to semi-analytically determine the outcomes of impacts between a more massive ``target'' object and a less massive ``projectile'' object. The outcome of a particular collision depends on the impact angle, impact velocity, and masses and radii of the colliding objects.

The largest object remaining after the collision is named the largest remnant and any less massive objects the collision produces are named fragments. During the course of a simulation, the module produces a collision report that records all collisions that occur during the runtime. Users can then input this file into the Differentiated Body Composition Tracker to study how the CMFs of objects change over the course of a simulation as a result of impacts (see Section \ref{sec:composition_tracking} for more details).

The fragmentation module contains eight possible outcomes for collisions, which we combine into six categories in this paper for brevity: bounces, mergers, erosions, accretions, and hit-and-runs. Bounces occur when the colliding object remains completely intact after impact. Mergers occur when the two colliding objects combine into a single object. Erosions lower the mass of the target and turn the projectile and the removed mass into equal-sized fragments. Accretions raise the mass of the target and turn the remaining mass in the projectile into equal-sized fragments. The minimum possible mass of these fragments is a variable that can be changed by the user. For our study, we choose this minimum mass to be half of the planetesimal mass (0.0193 $M_{\oplus}$). Finally, hit-and-runs are similar to accretions but allow the projectile to remain intact, albeit with less mass. For more information on the fragmentation module, see \citet{childs2022}. 

\subsection{Composition Tracking and Interior Modeling}
\label{sec:composition_tracking}

To track the changes in CMF produced by collisions between objects, \citet{Ferich2025} created a module called the Differentiated Body Composition Tracker (DBCT). This post-processing software takes a collision report from the fragmentation module for \reb\ and a composition input file that details the initial CMFs of all objects in a simulation and produces a composition output file that returns the CMFs of all objects remaining at the end of a simulation. The DBCT assumes that all objects are perfectly spherical and have two distinct layers of core and mantle material and determines the CMF changes from an individual impact by using the collision type, impact angle, and core and outer radii of the two colliding objects. Users can fine tune how much core material is transferred during a collision by setting the minimum and maximum fraction of material that can be included in the ejecta of an erosive or accretive collision; these parameters are known as $f_{min}$ and $f_{max}$ respectively.

For this paper, we set $f_{min} = 0.0$ and $f_{max} = 1.0$, which is the maximum possible range between the two. In addition, users can change the densities of the core and mantle material, which will determine the core and outer radii of the colliding objects. For this study, we set the density of core material to 7874.0 kg m$^{-3}$, the density of iron, and the density of mantle material to 3000.0 kg m$^{-3}$. For more details on the DBCT, see \citet{Ferich2025}.

\begin{figure}
	\includegraphics[width=\columnwidth]{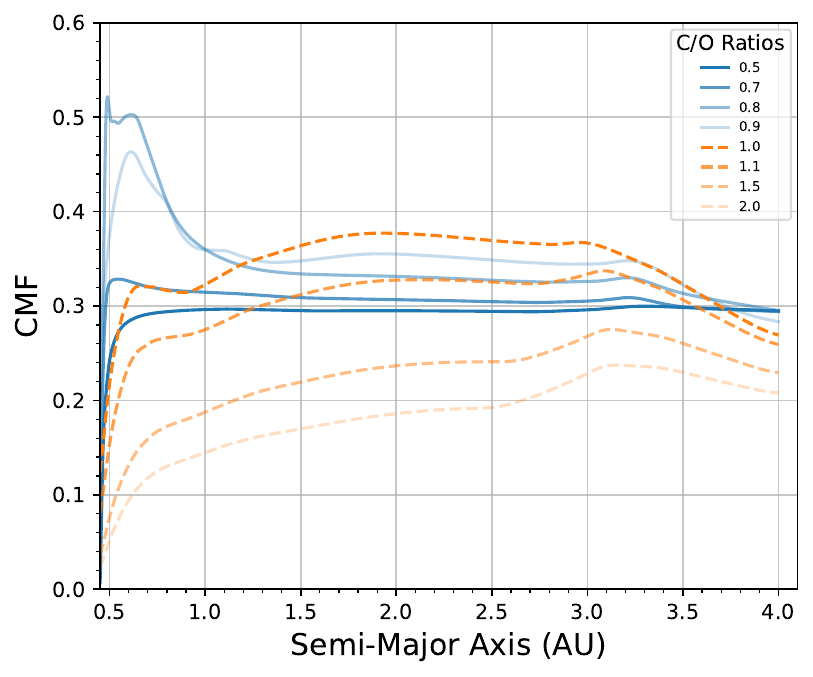}
    \caption{The effective CMF at different distances in a protoplanetary disk depending on the C/O ratio. CMF values at a particular location are calculated by finding the ratio between the sum of the surface densities of siderophile elements and the total surface density of the disk. Objects in our simulations are given a CMF depending on where their positions lie on these profiles.}
    \label{fig:initial_cmf_distributions}
\end{figure}

The results we obtained from \citet{Shakespeare2025} show that the C/O ratio of a protoplanetary disk can determine how much of a particular element condenses at certain regions within the disk. A portion of these condensed elements are siderophiles: elements that will settle into the core of a larger object. By adding the surface densities of these siderophiles together\footnote{Similar to Earth's core, the majority of the mass from these siderophiles comes from iron followed by nickel.} and dividing it by the disk's total surface density, we can create effective CMF distributions across the disk for each C/O ratio, which are shown in Figure \ref{fig:initial_cmf_distributions}. 

The plot indicates that the C/O ratio can significantly change the effective CMF at a particular location in the disk. This effect is mainly due to how the combined surface densities of carbon and oxygen, which are not considered siderophiles, vary between different C/O ratios. At a ratio 0.5, which is near the solar value, the majority of the disk will have a CMF of just under 0.3, a value close to the CMF of Earth ($\approx$0.33 \citep{Stacey2005}). As the C/O ratio increases, the CMF profile shifts upwards, with significant spikes below 1.0 AU for ratios of 0.8 and 0.9. This could lead to rocky planets with CMFs higher than that of Earth. For high C/O ratios (values of 1.0 and above), the shape of the CMF profile changes dramatically and begins to shift downwards with increasing C/O ratios, which could lead to rocky planets with CMFs below that of Earth. We use this plot to create composition input files for each C/O ratio by mapping their semi-major axes onto the corresponding profile and giving them the CMF found at that particular distance.

We use these files with the DBCT to determine whether or not the CMFs of planets depend on the C/O ratio of the system and whether or not these discrepancies are observable. To do so, we first find the average planetary CMF and mass for each C/O ratio. We then input these values into \magrathea\ \citep{Huang2022} --- a planetary interior modeling software that now has the ability to model carbon planets --- to find the average outer planetary radius for each C/O ratio. When using this code, we assume the interiors of both silicon and carbon planets have continuous temperature profiles and that all of them have a surface temperature of 1200 K and no significant hydrosphere or atmosphere. Once we have these results, we use the average planetary mass and average planetary radius to calculate the average bulk planetary density for each C/O ratios and compare them to see if there are significant differences between carbon and silicon planets.  We note that we do not consider any detailed effects of the oxidation state of the planet-forming material, which can cause variation in the core mass fraction of a forming planet \citep{Dale:2026}.

\section{Results}
\label{sec:results}

We present the results from our analysis of all 160 simulations in the following sections. In the first, we analyze the statistics from the final planetary systems to try any find any observable differences between the architectures of systems formed from disks with different C/O ratios. In the second section, we look at the percentage of siderophiles that compose each planet, which we consider to be a planet's CMF, and try to discern any correlation between the CMFs and bulk densities of planets and whether they are majority carbon or silicon. For all results, we assume that a planet is an object with a mass equal to or larger than the embryo mass, a fragment is an object with a mass equal to or below the planetesimal mass, and a planetesimal is any object with mass between the initial planetesimal and embryo masses. In addition, we consider C/O ratios with values below unity as low ratios and systems produced in those disks as silicate-dominant, and we will consider C/O ratios with values of unity or above as high ratios and systems produced in those disks as carbon-dominant (see Section \ref{sec:sequential_condensation_model} for reasoning).

\subsection{System Architectures}
\label{sec:system_architectures}
\subsubsection{Regular Disks}
\label{sec:system_architectures_regular_disks}

\begin{figure*}
	\includegraphics[width=\textwidth]{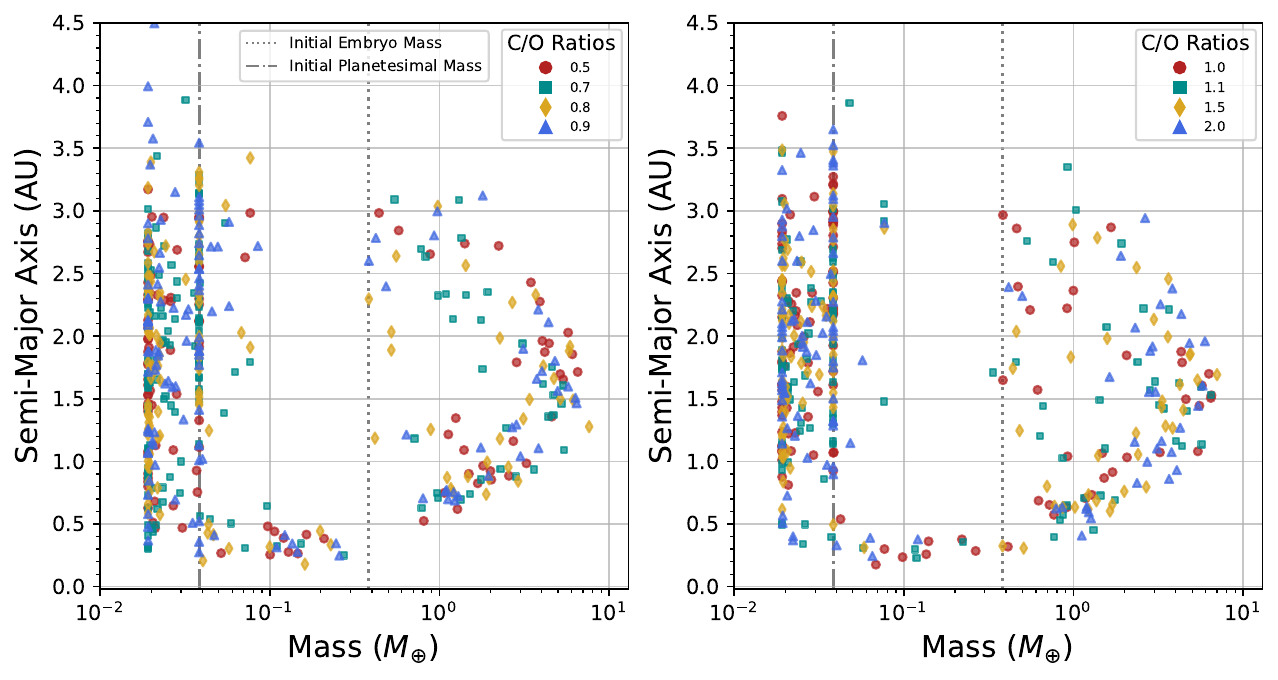}
    \caption{Final semi-major axes and masses for all objects produced in simulations with regular disks. The left panel shows objects produced in disks with low C/O ratios and the right one shows planets in high C/O ratio disks. The dash-dotted line shows the initial mass of planestesimals in our simulations while the dotted line shows the initial mass of planetary embryos. We consider planets to be anything at or above the initial embryo mass, fragments to be anything below the initial planetesimal mass, and planetesimals to be any object at or above the initial planetesimal mass but below the initial embryo mass.}
    \label{fig:regular_disk_orbits}
\end{figure*}

Fig. \ref{fig:regular_disk_orbits} contains the masses and final semi-major axes for all objects produced in our regular disk simulations. The left panel shows objects formed in disks with C/O ratios below 1.0 and the right panel shows objects formed in disks with C/O ratios above 1.0. The dotted line indicates the initial mass of planetary embryos in our simulations while the dashed line indicates the initial mass of planetesimals in our simulations.

For all C/O ratios, most planets lie between 0.3 and 3.5 AU with no easily discernible trends between the C/O ratio and the semi-major axis. There are two planets not shown on these plots: one with a mass of 0.460 $M_{\oplus}$, a semi-major axis of 10.4 AU, and an eccentricity of 0.147 that formed in a disk with a C/O ratio of 1.0 and one with a mass of 0.422 $M_{\oplus}$, a semi-major axis of 10.8 AU, and an eccentricity of 0.324 that formed in a disk with a C/O ratio of 2.0. Both of these objects are embryos that experienced a merging collision with a planetesimal that drastically changed their orbits. We consider these two planets outliers and do not include them when performing our statistical analysis on the architectures of these planetary systems.

In each panel of Fig. \ref{fig:regular_disk_orbits}, larger planets tend to form closer to the middle of the 0.3 to 3.5 AU range, giving these areas of the plots distinctive arrowhead shapes. It's likely that embryos in the middle of the disk have easier access to material from different parts of the disk, allowing them to grow to sizes larger than embryos found near the inner and outer edges of the initial disk. However, a number of smaller planets, especially ones formed in disks with high C/O ratios, still form at the central regions of the disk. Overall, we were able to create a wide variety of sub-earths, earth analogues, and Super-Earths with masses ranging from 0.385 $M_{\oplus}$ to 7.62 $M_{\oplus}$. 

We also produced a large number of fragments and planetesimals during these simulations. Most of these objects are contained within 0.3 AU and 3.5 AU like the planets, though there are some exceptions above and below this range. The majority of fragments have a mass of 0.0193 $M_{\oplus}$, which is half of the initial planetesimal mass and the minimum fragment mass for fragments produced by collisions in our simulations (see section \ref{sec:nbody_simulations_and_fragmentation} for more detail). A large number of planetesimals still have their initial mass of 0.0385 $M_{\oplus}$; these are likely objects that did not undergo any collisions during a particular simulation, though they could also be the results of fragment mergers. Planetesimals, especially ones with masses above 0.1 $M_{\oplus}$, are scarce for all C/O ratios, likely due to large numbers of merging collisions with planetary embryos. However, there is a notable group of planetesimals with large masses at small semi-major axes in both panels, which are fragments from disruptive collisions that underwent subsequent collisions to grow to their final masses.

\begin{table*}
    \begin{center}
    \caption{Averages and standard deviations for different system and planetary parameters that were produced by regular disks with different C/O ratios.}
    \label{tab:regular_planet_stats}
    \begin{tabular}{ccccccccc}
    \hline
         C/O Ratio & $N_{p}$ & M ($M_{\oplus}$) & $a_{min}$ (AU) & $a_{max}$ (AU) & e & inc (Radians) & $a_{\Delta}$ (AU)\\
         \hline
         0.5 & 3.20 $\pm$ 0.600 & 2.70 $\pm$ 1.78 & 0.776 $\pm$ 0.193 & 2.43 $\pm$ 0.405 & 0.0898 $\pm$ 0.0493 & 0.0729 $\pm$ 0.6001 & 0.753 $\pm$ 0.214\\
         0.7 & 3.30 $\pm$ 0.781 & 2.30 $\pm$ 1.60 & 0.752 $\pm$ 0.212 & 2.46 $\pm$ 0.421 & 0.106  $\pm$ 0.0733  & 0.0686 $\pm$ 0.0668 & 0.742 $\pm$ 0.277\\
         0.8 & 3.20 $\pm$ 0.748 & 2.47 $\pm$ 1.97 & 0.798 $\pm$ 0.248 & 2.18 $\pm$ 0.453 & 0.121  $\pm$ 0.136  & 0.0862 $\pm$ 0.0833 & 0.629 $\pm$ 0.219\\
         0.9 & 3.20 $\pm$ 0.748 & 2.41 $\pm$ 1.85 & 0.727 $\pm$ 0.254 & 2.39 $\pm$ 0.539 & 0.101  $\pm$ 0.086  & 0.0838 $\pm$ 0.0779 & 0.755 $\pm$ 0.308\\
         1.0 & 3.60 $\pm$ 1.11 & 2.12 $\pm$ 2.06 & 0.588 $\pm$ 0.191 & 2.21 $\pm$ 0.511 & 0.109  $\pm$ 0.0744  & 0.0989 $\pm$ 0.0970 & 0.648 $\pm$ 0.246\\
         1.1 & 3.70 $\pm$ 0.458 & 2.12 $\pm$ 1.67 & 0.570 $\pm$ 0.126 & 2.45 $\pm$ 0.493 & 0.123  $\pm$ 0.0889  & 0.112  $\pm$ 0.0867  & 0.697 $\pm$ 0.365\\
         1.5 & 3.80 $\pm$ 0.980 & 2.28 $\pm$ 1.63 & 0.590 $\pm$ 0.146 & 2.27 $\pm$ 0.414 & 0.123  $\pm$ 0.0766  & 0.115  $\pm$ 0.107 & 0.599 $\pm$ 0.262\\
         2.0 & 3.40 $\pm$ 0.663 & 2.70 $\pm$ 1.52 & 0.668 $\pm$ 0.150 & 2.22 $\pm$ 0.332 & 0.117  $\pm$ 0.0100  & 0.0732 $\pm$ 0.0656 & 0.673 $\pm$ 0.266\\
         \hline
    \end{tabular}
    \end{center}
\end{table*}

Table \ref{tab:regular_planet_stats} shows average values of different parameters for planets and planetary systems produced by the 10 regular disks with a particular C/O ratio. These parameters include the number of planets in a particular system, the masses of planets, the minimum semi-major axis and maximum semi-major axis that planets have in a system, the eccentricities and inclinations of planets, and the average distance between planets. 

On average, all C/O ratios produce three to four planets. The mean masses of planets fall between 2.12 $M_{\oplus}$ and 2.70 $M_{\oplus}$ with no clear difference between high and low C/O ratio systems. This uniformity comes from the total initial mass available in the disk, which we hold constant for every disk type (20 $M_{\oplus}$) in order to form planets in the Super-Earth mass range. If we remove this assumption, the discrepancies between the surface density profiles in the first panel of Figure \ref{fig:surface_density_plot} could lead to a correlation between the average planetary mass and C/O ratio of the system. We discuss this topic further in Section \ref{sec:future_work}.

The innermost planets in high C/O ratio systems tend to be closer to their host stars than ones in low C/O ratio systems. This result could be a reflection of the large surface densities that high C/O ratio disks have at short distances (see Figure \ref{fig:surface_density_plot}), which results in embryos and planetesimals that can form closer to the star. On average, the outermost planets tend to form with semi-major axes between 2 AU and 3 AU, a trend that appears unrelated to the C/O ratio of the disk. The eccentricites of planets seem to be greater for intermediate C/O ratios (0.9, 1.0, 1.1) and tend to decrease towards each end of the range. However, this could be a stochastic result, especially considering the fact that these eccentricities could decrease if the simulation time were increased. Inclinations are generally small and don't follow any clear correlation with C/O ratio. Finally, the average distances between planets in a given system do not seem to depend strongly on whether they are carbon-dominant or silicate-dominant.

\subsubsection{Compact Disks}
\label{sec:system_architectures_compact_disks}

\begin{figure*}
	\includegraphics[width=\textwidth]{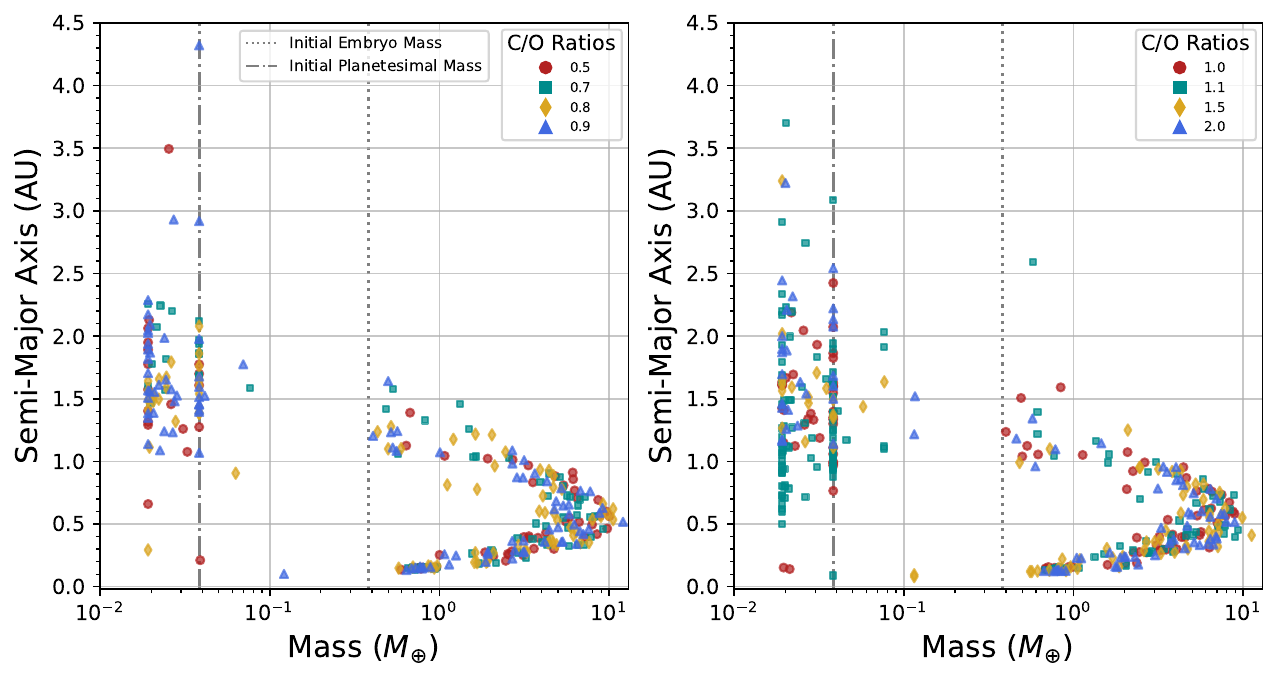}
    \caption{Same as Figure \ref{fig:regular_disk_orbits} but for objects produced in compact disks.}
    \label{fig:compact_disk_orbits}
\end{figure*}

Similar to Fig. \ref{fig:regular_disk_orbits}, Fig. \ref{fig:compact_disk_orbits} shows the masses and final semi-major axes for all objects produced in our compact disk simulations. The left panel shows objects formed in low C/O ratio disks and the right panel shows objects formed in high C/O ratio disks. The dotted line indicates the initial mass of planetary embryos in our simulations and the dashed line indicates the initial mass of planetesimals in our simulations.

Similar to regular disks, the orbits of planets produced by compact disks do not show any significant correlations with the system's C/O ratio. Most planets lie between 0.1 AU and 1.7 AU, with a single exception at around 2.6 AU. The masses of planets range from the initial embryo mass (0.385 $M_{\oplus}$) to around 11.2 $M_{\oplus}$, a value above the 10 $M_{\oplus}$ limit often attributed to Super-Earths \citep{Charbonneau2009} and around 4 $M_{\oplus}$ larger than the biggest planet formed in the regular disks.

The planetary mass difference between the two types of disks could be a result of multiple mechanisms. First, objects in the outer regions of the regular disks are more susceptible to the influence of the Jupiter-analogue in our simulations, leading to more ejections and less total mass available in the system. Second, Fig. \ref{fig:regular_disk_orbits} shows that the regular disks produced more fragments and contain more planetesimals than the compact disks, which indicates the objects in compact disks underwent more mergers and less disruptive collisions than regular disks; this leads to larger planets at the end of a given integration. 

This result might seem counterintuitive considering the higher orbital velocities found throughout the compact disk; however, in this case, the time at which these collisions occur is more important. Many of the collisions in the compact disk occur shortly after the start of a given simulation and before the Jupiter-analogue could perturb objects into more eccentric objects, which would lead to more violent collisions. In regular disks, collisions occur farther into simulations which allows the conditions for extreme impacts to develop. This discrepancy could be due to the shorter periods and closer orbits found in our compact disks. Shorter periods mean more orbits within a given amount of time while closer orbits will raise the relative probability of a collision within a single orbit. Combined, these two factors will lead to more collisions on shorter timescales in a simulation.

Fragments and planetesimals tend to reside farther away from the star than planets, with most of them having a semi-major axis between 0.5 AU and 2.0 AU. The compact disks produced fewer fragments and planetesimals compared to the regular disks for the same reason they formed planets with larger masses: the nature of the disk leads to more merger collisions than erosive ones. 

\begin{table*}
    \begin{center}
    \caption{Same as Table \ref{tab:regular_planet_stats} but for compact disks.}
    \label{tab:compact_planet_stats}
    \begin{tabular}{cccccccc}
    \hline
    \hline
         C/O Ratio & $N_{p}$ & M ($M_{\oplus}$) & $a_{min}$ (AU) & $a_{max}$ (AU) & e & inc & $a_{\Delta}$ (AU)\\
         \hline
         0.5 & 4.90 $\pm$ 0.700 & 3.94 $\pm$ 2.81 & 0.154 $\pm$ 0.0182 & 1.00 $\pm$ 0.157   & 0.0568 $\pm$ 0.0604 & 0.0472 $\pm$ 0.0380 & 0.218 $\pm$ 0.124\\
         0.7 & 5.10 $\pm$ 0.831 & 3.78 $\pm$ 2.52 & 0.166 $\pm$ 0.0258 & 1.19 $\pm$ 0.235   & 0.0714 $\pm$ 0.0540 & 0.0493 $\pm$ 0.0391 & 0.250 $\pm$ 0.147\\
         0.8 & 5.10 $\pm$ 0.700 & 3.78 $\pm$ 2.81 & 0.173 $\pm$ 0.0199 & 1.12 $\pm$ 0.124   & 0.0655 $\pm$ 0.0670 & 0.0437 $\pm$ 0.0421 & 0.231 $\pm$ 0.102\\
         0.9 & 5.40 $\pm$ 0.800 & 3.57 $\pm$ 2.68 & 0.164 $\pm$ 0.0263 & 1.15 $\pm$ 0.196   & 0.0612 $\pm$ 0.0627 & 0.0482 $\pm$ 0.0419 & 0.224 $\pm$ 0.115\\
         1.0 & 5.40 $\pm$ 0.800 & 3.57 $\pm$ 2.46 & 0.160 $\pm$ 0.0169 & 1.14 $\pm$ 0.224   & 0.0586 $\pm$ 0.0398 & 0.0470 $\pm$ 0.0374 & 0.224 $\pm$ 0.147\\
         1.1 & 5.30 $\pm$ 0.458 & 3.62 $\pm$ 2.61 & 0.145 $\pm$ 0.0107 & 1.19 $\pm$ 0.490   & 0.0783 $\pm$ 0.0763 & 0.0477 $\pm$ 0.0472 & 0.243 $\pm$ 0.212\\
         1.5 & 5.10 $\pm$ 0.700 & 3.81 $\pm$ 2.66 & 0.137 $\pm$ 0.0136 & 0.966 $\pm$ 0.121 & 0.0600 $\pm$ 0.0533 & 0.0401 $\pm$ 0.0291 & 0.202 $\pm$ 0.100\\
         2.0 & 5.30 $\pm$ 0.781 & 3.65 $\pm$ 2.38 & 0.139 $\pm$ 0.0180 & 1.02 $\pm$ 0.160   & 0.0557 $\pm$ 0.0368 & 0.0474 $\pm$ 0.0365 & 0.205 $\pm$ 0.104\\
         \hline
    \end{tabular}
    \end{center}
\end{table*}

Table \ref{tab:compact_planet_stats} has the same format as Table \ref{tab:regular_planet_stats} but now displays the statistics for planets and systems produced by compact disks. Similar to the regular disks, these results suggest that low C/O ratio systems often produce less planets overall than high C/O ones. However, the correlation here is less significant. Overall, compact disks formed around two more planets on average than regular disks, which could be a result of more mergers and less ejections. They also formed larger planets on average for the same reasons. These results also show that some of the smallest C/O ratios form some of the larger planets, but the correlation is weak. 

With average semi-major axes ranging from 0.37 AU to 1.19 AU, compact disk planets form at smaller distances from their host star than regular disk planets. Similar to regular disks, the statistics for compact disks show that high C/O ratio innermost planets tend to form slightly closer towards the host star than low C/O ratio ones, which is again a result of the higher surface densities at shorter distance for higher C/O ratios. The maximum semi-major axes, eccentricities, inclinations, and planet separations do not appear to have any noticeable correlation with C/O ratios for planets formed in compact disks.

\subsubsection{Gap Complexity}
\label{sec:gap_complexity}

Gap complexity ($C$) is a unitless metric that determines how evenly spaced planets are in systems of three or more planets. This measure of complexity was described in \citet{Lopez_Ruiz1995} and first applied to study the architectures of planetary systems by \citet{Gilbert2020}. We apply it to the results from our regular and compact disk simulations to determine if it could be used to differentiate between a system of silicon planets and a system of carbon planets. 

The equation for gap complexity is given by the following:
\begin{equation}
    C = -K\left(\sum_{i=1}^{n-1}p_{i}^\star\ln(p_i^\star)\right)\left(\sum_{i=1}^{n-1}\left(p_i^\star - \frac{1}{n-1}\right)^2\right),
    \label{eq:gap_complexity}
\end{equation}
\begin{equation}
    p_i^\star = \frac{ln(P_{i+1}/P_i)}{ln(P_n/P_1)},
    \label{eq:p_i_star}
\end{equation}
where n is the number of planets in the system, $P_{i+1}/P_i$ is the period ratio between adjacent planets, and $P_{n}/P_1$ is the period ratio between the outermost and innermost planets. $K$ is a normalization factor that depends on the number of planets in the system. It is equal to $1/C_{max}$, where $C_{max}$ = [0.105, 0.212, 0.291, 0.350, 0.398, 0.437] for systems of three to eight planets respectively.

\begin{table}[ht]
    \centering
    \caption{Average gap complexities and standard deviations for the systems formed by each C/O ratio in both our regular and compact disks.}
    \label{tab:avg_gap_complexities}
    \begin{tabular}{ccc}
    \hline
    \hline
         C/O Ratio & $C_{regular}$ & $C_{compact}$\\
         \hline
         0.5 & 0.142 $\pm$ 0.138 & 0.0535 $\pm$ 0.0459\\
         0.7 & 0.217 $\pm$ 0.141 & 0.0838 $\pm$ 0.0400\\
         0.8 & 0.200 $\pm$ 0.296 & 0.0675 $\pm$ 0.0377\\
         0.9 & 0.140 $\pm$ 0.207 & 0.0626 $\pm$ 0.0236\\
         1.0 & 0.250 $\pm$ 0.172 & 0.0818 $\pm$ 0.0579\\
         1.1 & 0.321 $\pm$ 0.237 & 0.0857 $\pm$ 0.0453\\
         1.5 & 0.216 $\pm$ 0.178 & 0.0987 $\pm$ 0.0451\\
         2.0 & 0.253 $\pm$ 0.188 & 0.0614 $\pm$ 0.0298\\
         \hline
    \end{tabular}
\end{table}

We obtain gap complexity values for every system with over three planets that we produced in our simulations. Table \ref{tab:avg_gap_complexities} lists the average gap complexities and their standard deviations for each C/O ratio in both our regular and compact disks. For reference, the gap complexity of the Solar System is around 0.126 \citep{Rice2024}. Overall, the gap complexities of systems formed in our regular disks are higher than the ones formed from our compact disks, likely due to the greater extent of and the larger number of disruptive collisions that occur during planet formation in this type of disk. However, the high standard deviations indicate that individual systems can deviate far from the mean. This variance is partially due to the cutoff we chose for what constitutes a planet in these simulations (having a mass greater or equal to the mass of a planetary embryo) and partially due to the small sample size for each C/O ratio (below 10 in most cases).

\begin{figure}
	\includegraphics[width=\columnwidth]{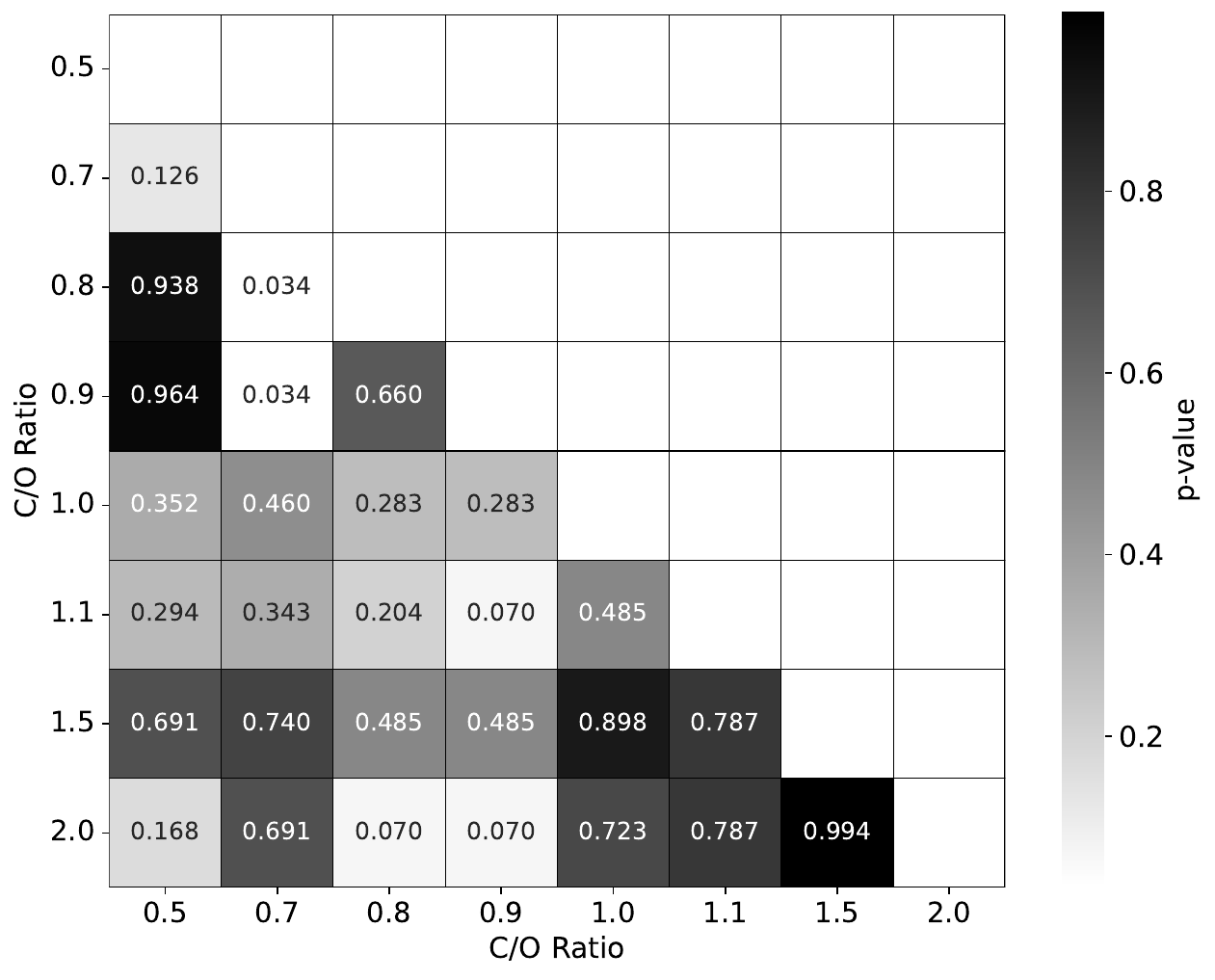}
    \caption{p-values obtained from our KS tests on the gap complexities produced by systems with different C/O ratios. These systems were produced by our regular disk simulations.}
    \label{fig:ks_tests_regular_gap_complexity}\end{figure}

\begin{figure}
	\includegraphics[width=\columnwidth]{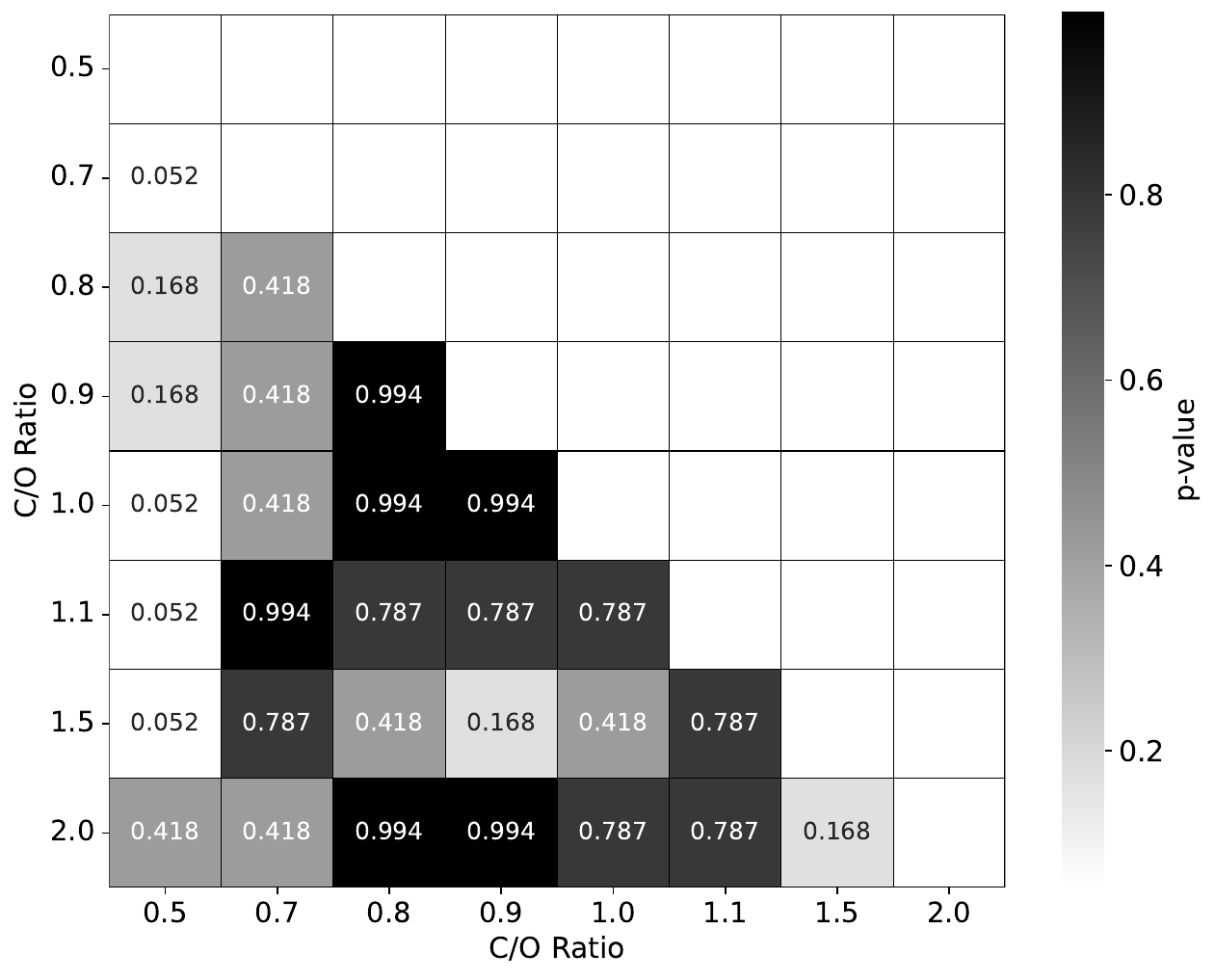}
    \caption{Same as Figure \ref{fig:ks_tests_regular_gap_complexity} but for systems produced by our compact disk simulations.}
    \label{fig:ks_tests_compact_gap_complexity}\end{figure}
    
We perform two-sample Kolmogorov-Smirnov (KS) and k-sample Anderson-Darling (AD) tests on the gap complexities produced by different C/O ratios to determine if this metric is a reliable way to distinguish between systems with silicon and carbon planets. Lower p-values indicate that the gap complexities more likely come from separate distributions that are determined by their C/O ratio. If the gap complexities between silicon-dominated and carbon-dominated systems are different, then we would expect to see lower p-values when comparing low ratios to high ratios and higher p-values when we compare low ratios with each other and high ratios with each other. 

Our KS and AD tests yield similar trends, so we have chosen to only display the results from our KS tests. Figure \ref{fig:ks_tests_regular_gap_complexity} shows the p-values from these tests for systems that come from our regular disk simulations while Figure \ref{fig:ks_tests_compact_gap_complexity} shows the p-values from these tests for systems that come from our compact disk simulations. For our regular disks, the p-values when comparing low and high ratios seem to be generally lower than the p-values comparing low to low and high to high. However, the plot shows plenty of counterexamples to this trend, which is nonexistent in the plot for our compact disks. Instead, Figure \ref{fig:ks_tests_compact_gap_complexity} shows that the gap complexities for systems with a C/O ratio of 0.5 are significantly different than the ones from other systems. Additionally, only two of our p-values were below the often used threshold of 0.05 for rejecting the null hypothesis. These values both come from Figure \ref{fig:ks_tests_regular_gap_complexity} and both come from KS tests performed between two low C/O ratios.

With our results, we conclude that gap complexities are not a reliable way to distinguish between silicon-dominant systems and carbon-dominant systems. Gap complexities can vary significantly for systems with similar C/O ratios and are not significantly different for systems with low and high ratios. We note that these results could partially be due to our small sample sizes for each ratio ($\leq$ 10) and could change with more data.

\subsection{Planet Compositions}
\label{sec:planet_compositions}

In this section, we use the DBCT to examine if the CMFs of planets depend on the C/O ratio of the disk in which they form. We have created two types of CMF distributions for both regular and compact disks: one where every object in the simulation begins with a CMF of 0.3 and one where the CMF of each object is determined by its semi-major axis and the C/O ratio (see Figure \ref{fig:initial_cmf_distributions}). The first distribution, which we refer to as the uniform distribution, will show how important collisional history is for determining the CMFs of planets in these simulations. The second distribution, which we call the non-uniform distribution, will show how important the C/O ratio of a disk is in determining the final CMFs of planets.

\subsubsection{Regular Disks}
\label{sec:planet_compositions_regular_disks}

\begin{figure}
	\includegraphics[width=\columnwidth]{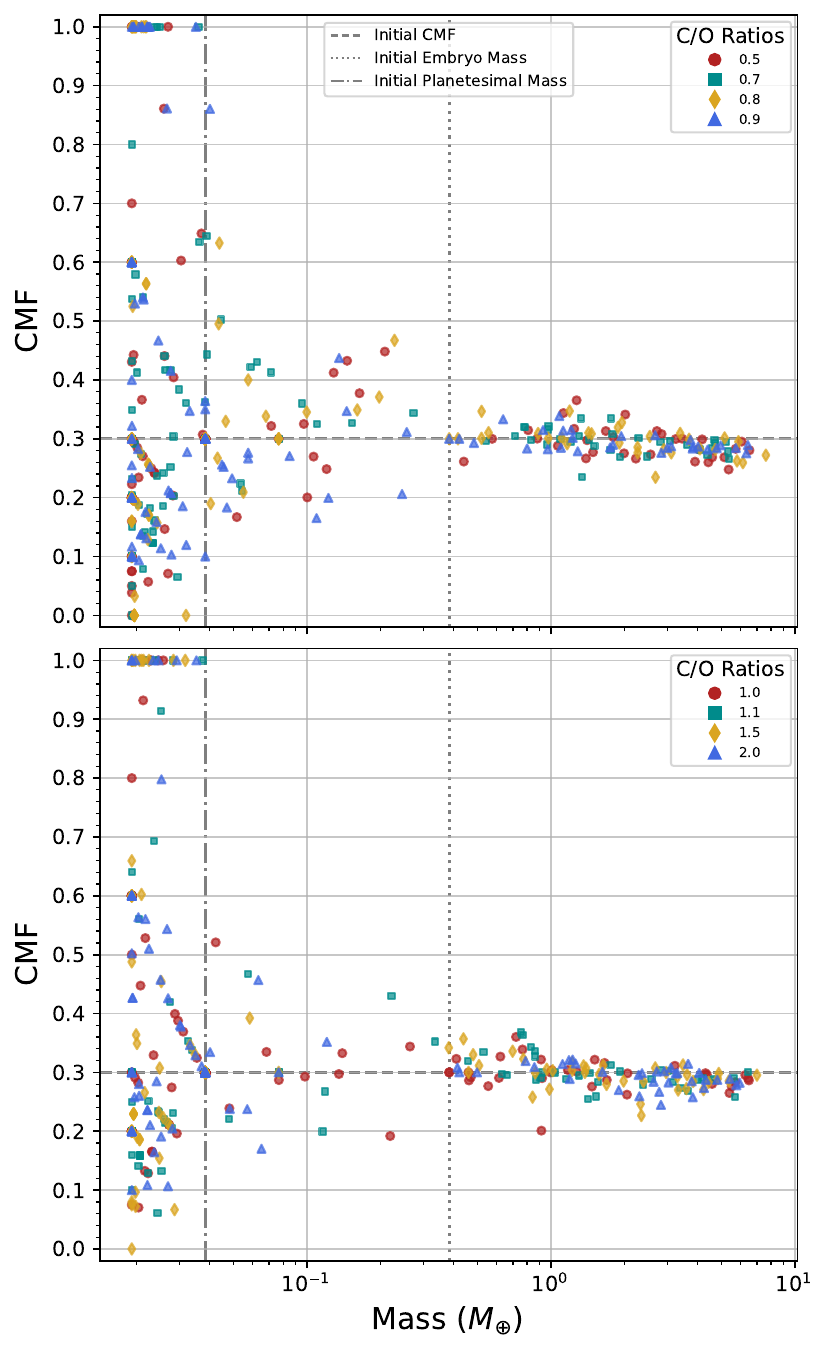}
    \caption{Masses and CMFs of objects produced in regular disks with different C/O ratios. For this plot, we assume that all planetary embryos and planetesimals begin with a uniform CMF of 0.3 at the start of every simulation. The top panel contains objects produced in disks with low C/O ratios while the bottom panel contains objects from disks with high C/O ratios.}
    \label{fig:uni_regular_cmfs}
\end{figure}

Figure \ref{fig:uni_regular_cmfs} shows the CMFs of all planets, planetesimals, and fragments produced by all 80 regular disk simulations with uniform initial CMFs. The panel on the top contains objects produced by low C/O ratio disks, and the panel on the bottom contains objects produced by high C/O ratio disks with the specific color and shape of each point denoting the exact C/O ratio. The dotted line indicates the initial mass of embryos in our simulation, the dash-dotted line indicates the initial mass of planetesimals in our simulations, and the dashed line shows the CMFs initially given to all objects the start of a simulation, which is 0.3. Similar to Section \ref{sec:system_architectures}, we consider planets to be any object with a mass with or above the initial embryo mass (0.385 $M_{\oplus}$), fragments to be any object below the planetesimal mass (0.0385 $M_{\oplus}$), and planetesimals to be any objects in between these two values or still possessing the initial planetesimal mass.

The results in the figure reflect what was found for uniform disks in \citet{Ferich2025}: the vast majority of planets obtain CMFs that are close to the initial CMF, while fragments and planetesimals have more diverse compositions. This discrepancy is due to the amount of mass these different objects must accrete to reach their respective sizes. For example, a planetary embryo could undergo one or multiple extreme erosive collisions early in its history that remove most of its mantle. However, it will then accrete a large amount of material from the disk to eventually become a planet, removing the composition-changing effects of any early disruptive collisions and leaving it with a final CMF close to the average initial CMF of the disk.

From this result, we can conclude that if energetic collisions are what typically produce a planet with an extreme CMF, then they must occur late in that planet's formation history. All of our planets have CMFs between 0.2 and 0.4, indicating that the collisional histories of planets are similar between C/O ratios and do not play a large role in affecting their final CMFs.

\begin{figure}
	\includegraphics[width=\columnwidth]{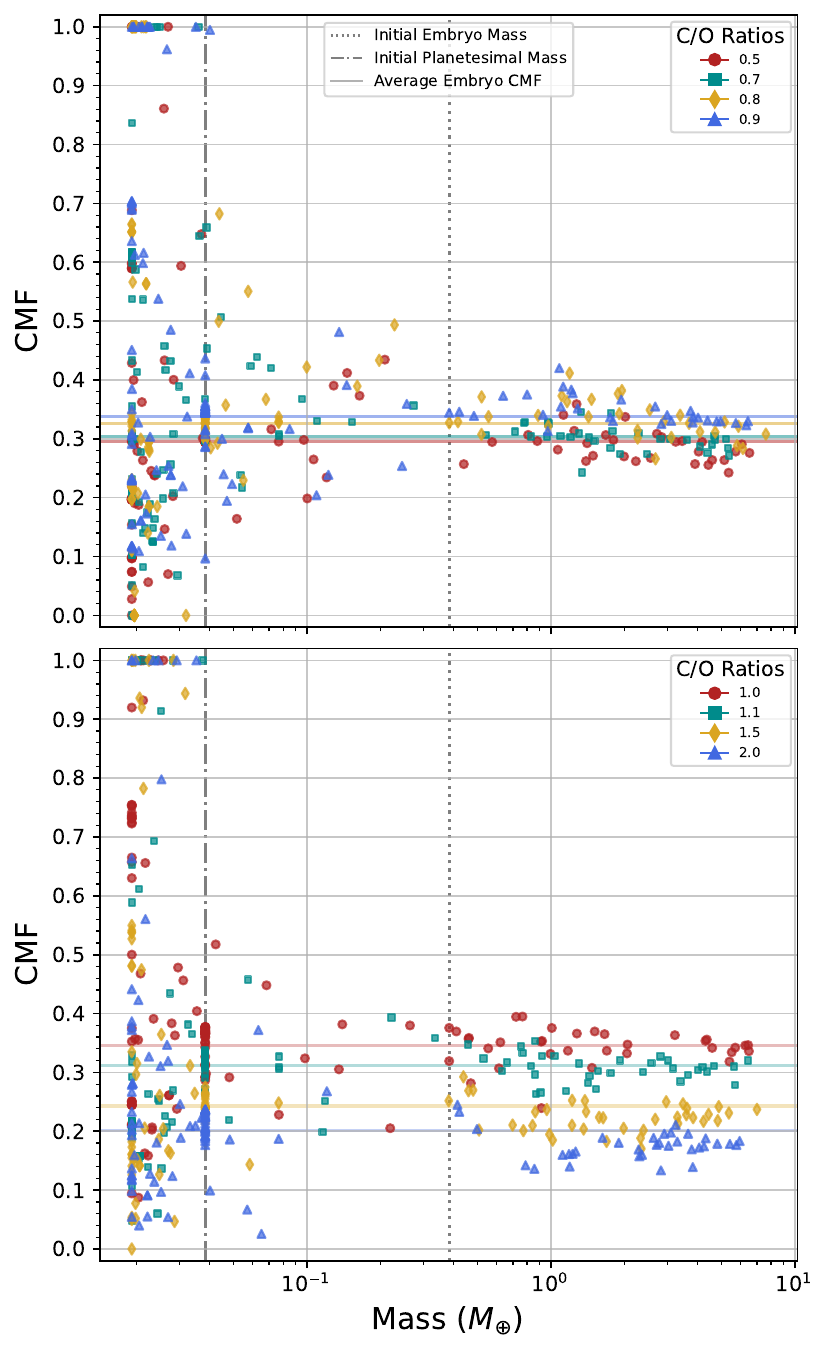}
    \caption{Masses and CMFs of objects produced in regular disks with different C/O ratios. For this plot, we assume that planetary embryos and planetesimals begin each simulation with a CMF that depends on the disk's C/O ratio and their location within the disk. Figure \ref{fig:initial_cmf_distributions} shows the profiles used to produce these non-uniform distributions of core material. The top panel contains objects produced in disks with low C/O ratios while the bottom panel contains objects from disks with high C/O ratios. Horizontal lines show the average embryo CMF at the start of each simulation for each C/O ratio.}
    \label{fig:nu_regular_cmfs}
\end{figure}

\begin{table}
    \centering
    \caption{Average planetary embryo CMF for each C/O ratio.}
    \label{tab:initial_average_embryo_cmfs}
    \begin{tabular}{cc}
    \hline
    \hline
         C/O Ratio & CMF\\
         \hline
         0.5 & 0.296\\
         0.7 & 0.304\\
         0.8 & 0.326\\
         0.9 & 0.338\\
         1.0 & 0.345\\
         1.1 & 0.312\\
         1.5 & 0.243\\
         2.0 & 0.201\\
         \hline
    \end{tabular}
    \tablecomments{These means apply to simulations using both regular and compact disks.}
\end{table}

Figure \ref{fig:nu_regular_cmfs} shows the same objects as Figure \ref{fig:uni_regular_cmfs}. However, instead of assuming the embryos and planetesimals begin with constant CMFs, we now assume they start with CMFs determined by our non-uniform distributions of initial CMFs. Figure \ref{fig:nu_regular_cmfs} also includes lines that indicate the average embryo CMF for each C/O ratio (the exact values are shown in Table \ref{tab:initial_average_embryo_cmfs}). The plot shows that most planets tend to remain close to the initial average embryo CMF for a particular C/O ratio, similar to the results found for the uniform distribution. The variations from this average could be due to disruptive collisions or accreting fragments or planetesimals with different CMFs from distant regions in the disk. It's also possible an embryo only accreted material with a similar CMF on its way to becoming a planet, leaving it with a CMF close to its original value.

\begin{table}
    \centering
    \caption{The average, minimum, and maximum planetary CMF values for each C/O ratio in our regular disk simulations.}
    \label{tab:regular_cmfs}
    \begin{tabular}{cccc}
    \hline
    \hline
         C/O Ratio & Avg. CMF & Min CMF & Max CMF\\
         \hline
         0.5 & 0.288 $\pm$ 0.0263 & 0.242 & 0.359 \\
         0.7 & 0.302 $\pm$ 0.0211 & 0.243 & 0.346 \\
         0.8 & 0.332 $\pm$ 0.0321 & 0.266 & 0.411 \\
         0.9 & 0.347 $\pm$ 0.0239 & 0.313 & 0.420 \\
         1.0 & 0.344 $\pm$ 0.0298 & 0.239 & 0.395 \\
         1.1 & 0.307 $\pm$ 0.0221 & 0.263 & 0.353 \\
         1.5 & 0.223 $\pm$ 0.0255 & 0.170 & 0.292 \\
         2.0 & 0.177 $\pm$ 0.0246 & 0.134 & 0.245 \\
         \hline
    \end{tabular}
\end{table}

Table \ref{tab:regular_cmfs} shows the average, minimum, and maximum planetary CMF for each C/O ratio in our regular disk simulations. The highest planetary CMF we produced was 0.420 while the lowest value was 0.134. On average, the C/O ratios around the transition region between silicon and carbon planets (0.9 and 1.0) produced the largest planetary CMFs. Moving outwards from this region in both directions, the average planetary CMF decreases with the minimum value being 0.177 from a C/O ratio of 2.0. The CMF values produced by C/O ratios of 1.5 and 2.0 are significantly lower than the other six ratios, though as mentioned previously, C/O ratios this high have not been observed yet. When it comes to more realistic values (0.5 to 1.1), the average planetary CMF differs by 0.05 at most. However, these CMF discrepancies will play a significant role in determining the average bulk planetary densities we find with \magrathea. 

\begin{table}
    \begin{center}
    \caption{Typical or average planet for a system with a particular C/O ratio in our regular disk simulations.}
    \label{tab:regular_magrathea_planets}
    \begin{tabular}{ccccc}
    \hline
    \hline
         C/O Ratio & M ($M_{\oplus}$) & CMF & Radius ($R_{\oplus}$) & $\rho$ (g/$cm^{3}$)\\
         \hline
         0.5 & 2.70 & 0.288 & 1.32 & 6.44\\
         0.7 & 2.30 & 0.302 & 1.26 & 6.32\\
         0.8 & 2.47 & 0.332 & 1.28 & 6.56\\
         0.9 & 2.41 & 0.347 & 1.26 & 6.61\\
         1.0 & 2.12 & 0.344 & 1.31 & 5.17\\
         1.1 & 2.12 & 0.307 & 1.33 & 4.99\\
         1.5 & 2.28 & 0.223 & 1.39 & 4.70\\
         2.0 & 2.70 & 0.177 & 1.48 & 4.56\\
         \hline
    \end{tabular}
    \end{center}
    \tablecomments{The first four planets are silicate-dominated while the last four are carbon-dominated.}
\end{table}

With the results for the non-uniform distribution, we can now use \magrathea\ to model a typical planet for each C/O ratio by inputting the average planetary CMF (shown in Table \ref{tab:regular_magrathea_planets}), the average planetary mass, and whether or not the planet is silicate-dominated or carbon-dominated. Table \ref{tab:regular_magrathea_planets} shows the outer radius the model produces and the bulk density for each planet we calculated assuming each planet is spherical. 

Our results indicate that carbon planets typically have lower bulk densities than silicon planets. For the two highest C/O ratios, this result could be partially due to the small cores these planets typically have. However, the two other carbon planets we modeled (1.0 and 1.1) have CMFs that are comparable to the values we obtained for silicon planets, yet they have significantly smaller bulk densities, implying that the differences in the internal structure of these two types of planets plays an important role in determining their radii and bulk densities. Importantly, we find that a typical silicon planet produced by our regular disks can be up to 45\% more dense than carbon planets, a discrepancy that could be significant enough to observationally distinguish silicon and carbon planets. This result agrees with the claim in \citet{Royer2026} that a carbon planet with the same radius as a silicon planet would need to have a higher observed bulk density than its silicon counterpart. However, we discuss possible caveats to using this method to distinguish planet type in Section \ref{sec:distinguishing_carbon_rich_planet}.

\subsubsection{Compact Disks}
\label{sec:planet_compositions_compact_disks}

\begin{figure}
	\includegraphics[width=\columnwidth]{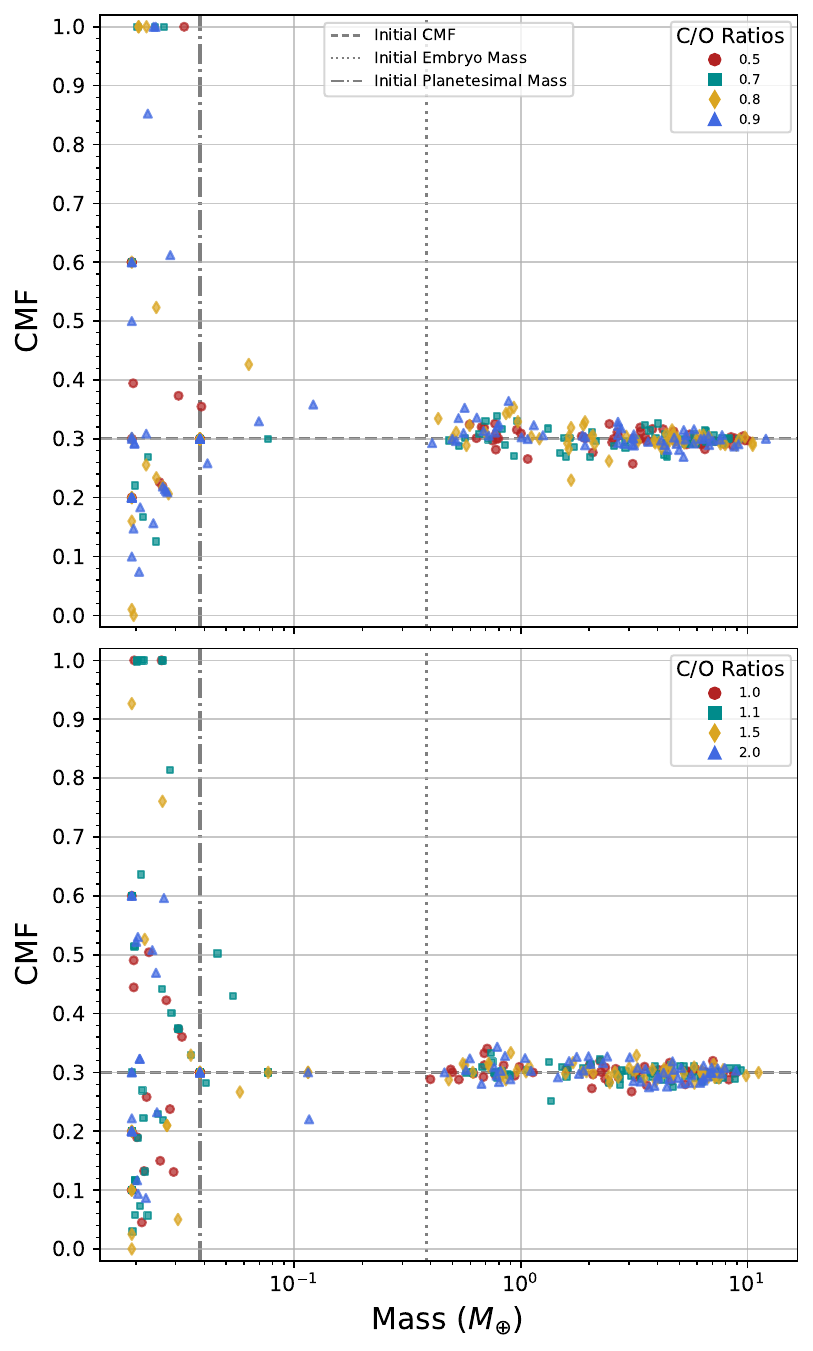}
    \caption{Same as Figure \ref{fig:uni_regular_cmfs} but for objects produced in our compact disks.}
    \label{fig:uni_compact_cmfs}
\end{figure}

Figure \ref{fig:uni_compact_cmfs} is the same as as Figure \ref{fig:uni_regular_cmfs} except it now shows objects that were produced in simulations with compact disks, which initially have the same number of objects as regular disks but extend from 0.09 AU to 0.8 AU instead of 0.3 AU to 4.0 AU. The planets in compact disks show even less variation from the initial CMF of 0.3 than the planets in the regular disk. Similar to our findings in \ref{sec:system_architectures}, these results indicate that most of the collisions during the course of these simulations were mergers, even more so than for the regular disk simulations. It also indicates that the distribution of CMFs in the non-uniform disk plays an important role in determining the final CMFs of planets.

\begin{figure}
	\includegraphics[width=\columnwidth]{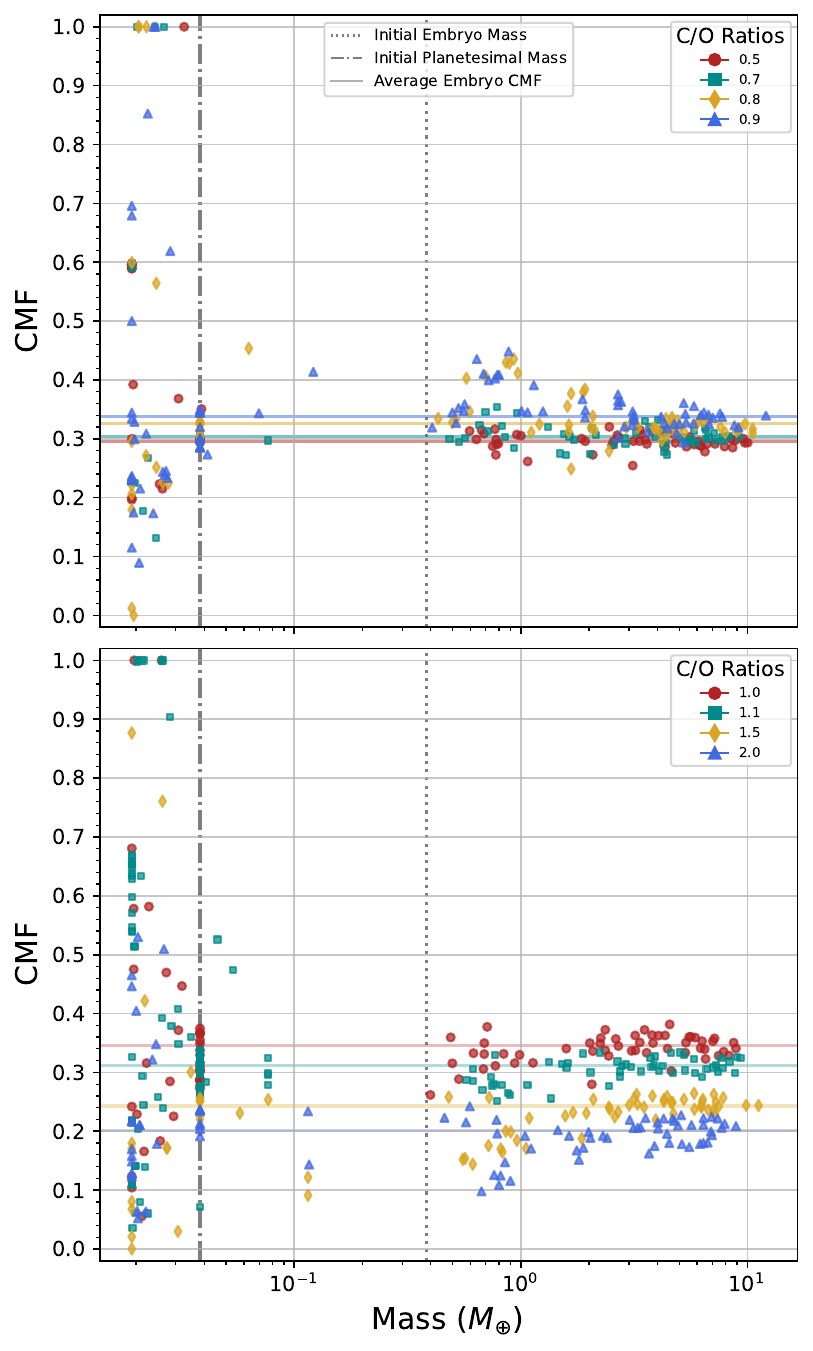}
    \caption{Same as Figure \ref{fig:nu_regular_cmfs} but for objects produced in our compact disks.}
    \label{fig:nu_compact_cmfs}
\end{figure}

Figure \ref{fig:nu_compact_cmfs} shows the CMFs of planets formed in compact disks where the initial CMFs are determined by our non-uniform distributions of core material. Similar to the results for regular disks, most planets lie close to the average embryo CMF for their respective C/O ratio, perhaps even more so than the planets formed in regular disks. However, there are notable exceptions for smaller planets with C/O values of 0.8, 0.9, 1.5, and 2.0. These groups of planets likely formed from embryos that did not accrete material from a wide range of the disk over the course of a simulation, leaving them with CMFs close to their initial values. 

\begin{table}
    \centering
    \caption{Same as Table \ref{tab:regular_cmfs} but for our compact disk results.}
    \label{tab:compact_cmfs}
    \begin{tabular}{ccccc}
    \hline
    \hline
         C/O Ratio & Avg. CMF & Min CMF & Max CMF\\
         \hline
         0.5 & 0.296 $\pm$ 0.0132 & 0.254 & 0.320\\
         0.7 & 0.305 $\pm$ 0.0174 & 0.272 & 0.355\\
         0.8 & 0.335 $\pm$ 0.0374 & 0.249 & 0.436\\
         0.9 & 0.348 $\pm$ 0.0320 & 0.294 & 0.448\\
         1.0 & 0.340 $\pm$ 0.0234 & 0.262 & 0.382\\
         1.1 & 0.304 $\pm$ 0.0206 & 0.251 & 0.334\\
         1.5 & 0.229 $\pm$ 0.0324 & 0.144 & 0.264\\
         2.0 & 0.191 $\pm$ 0.0320 & 0.0978& 0.242\\
         \hline
    \end{tabular}
\end{table}

Table \ref{tab:compact_cmfs} shows the average, minimum, and maximum planetary CMF for each C/O ratio in our compact disk simulations. The highest planetary CMF we produced was 0.448 while the lowest CMF was 0.0978 --- more extreme values than what we found for regular disks. However, when it comes to average planetary CMFs, the results are the same: the transition region (C/O ratios of 0.9 and 1.0) produced the highest average CMFs, the largest C/O ratios (1.5 and 2.0) produced the smallest average CMFs, and the more realistic C/O ratios (0.5 to 1.1) had average CMFs all within a range of about 0.05. Once again, we note that these CMF differences will affect the bulk densities we find using \magrathea.

\begin{table}
    \centering
    \caption{Same as Table \ref{tab:regular_magrathea_planets} but for the compact disk results.}
    \label{tab:compact_magrathea_planets}
    \begin{tabular}{ccccc}
    \hline
    \hline
         C/O Ratio & M ($M_{\oplus}$) & CMF & Radius ($R_{\oplus}$) & $\rho$ (g/$cm^3$)\\
         \hline
         0.5 & 3.94 & 0.296 & 1.46 & 6.99\\
         0.7 & 3.78 & 0.305 & 1.44 & 6.98\\
         0.8 & 3.78 & 0.335 & 1.43 & 7.15\\
         0.9 & 3.57 & 0.348 & 1.40 & 7.14\\
         1.0 & 3.57 & 0.340 & 1.53 & 5.52\\
         1.1 & 3.62 & 0.304 & 1.55 & 5.33\\
         1.5 & 3.81 & 0.229 & 1.58 & 5.34\\
         2.0 & 3.65 & 0.191 & 1.61 & 4.80\\
         \hline
    \end{tabular}
    \tablecomments{The first four planets are silicate-dominated while the last four are carbon-dominated.}
\end{table}

Using the same method described in Section \ref{sec:planet_compositions_regular_disks}, we input our simulation and composition tracking data for the compact disks into \magrathea\ to produce the properties of a typical planet for each C/O ratio, which can be found in Table \ref{tab:compact_magrathea_planets}. The results mirror what was found for regular disks: on average, carbon planets have lower bulk densities than silicate-rich ones due to the differences in their interior structures. This discrepancy is more pronounced for compact disks, where the difference in densities between the two types of planets can be as great as 79 percent --- a difference that could be observable. However, in the following section, we will discuss the possible limits and uncertainties to using bulk density to determine planet type.

\section{Discussion}
\label{sec:discussion}
\subsection{Distinguishing Carbon-rich Planets}
\label{sec:distinguishing_carbon_rich_planet}

Our results indicate that the bulk density of a planet could be used to determine if a planet is silicon-rich or carbon-rich. However, our interior models assumed that planets were made of only rocky material and contained no significant hydrospheres or atmospheres, even though planets above 5$M_{\oplus}$ are likely to retain an H/He envelope after formation \citep{Rogers2015}. Many of the planets we formed, especially the most massive ones ($\sim$10$M_{\oplus}$), could contain substantial layers of volatiles, which will affect their observed radii and bulk densities. We would require observations of any potential atmospheres surrounding these planets to have confidence in these models. Obtaining interior models that do include these additional layers would also provide a similar benefit.

Additionally, the measured values of exoplanetary masses and radii often have uncertainties that could make it difficult to confirm whether or not a planet is majority silicate or carbon with bulk density alone. To quantify this error, we derive how uncertainties in radius and mass measurement will propagate to uncertainty in a bulk density calculation. We find that a mass measurement with a precision of $\pm$0.01 $M_{\oplus}$ and a radius measurement with a precision of $\pm$0.01 $R_{\oplus}$ for an exoplanet with the same mass and radius as Earth would lead to an uncertainty of approximately $\pm$0.174 $g/cm^3$ for the calculated density. This error is around an order of magnitude lower than the highest difference in density we found between silicon and carbon planets (2.34 $g/cm^3$).

Finally, an exoplanet's CMF could lead to degeneracies in bulk density that could make distinguishing between silicon and carbon exoplanets difficult. A carbon planet with a particularly high CMF could be misidentified as a silicon planet, while a silicon planet with a particularly low CMF could be misidentified as a carbon planet. For example, results from \citet{Royer2026} show that a carbon planet with a mass of 1 $M_{\oplus}$ and CMF of approximately 0.5 would be similar in mass and radius to Earth, and therefore, have the same bulk density to an outside observer. While giant impacts during planet formation could lead to the abnormally high or low CMFs for this to occur, our results show that the majority of planets will have CMFs similar to the system average. In addition, our results indicate that carbon planets often have lower CMFs on average than silicon planets, so a high CMF carbon planet and a low CMF silicon planet would require a rare formation pathway. Therefore, we argue that CMF will not have a significant impact on how well bulk density can be used as a tracer of carbon planets.

Overall, we argue these limitations and caveats do not prevent bulk density from being a possible method to differentiate between carbon and silicon planets. However, more detailed interior models and better observations are necessary to make it a more reliable metric. Currently, bulk density on its own may not be enough to determine a planet's type with a high degree of certainty. However, in combination with follow up measurements of the stellar C/O ratio, it can be used as a positive indicator that an exoplanet has a carbon-rich composition. Additionally, if an exoplanet has a short enough period, then direct confirmation could come from detections of mantle material evaporating from its surface.

\subsection{Future Work}
\label{sec:future_work}

\begin{table}
    \centering
    \caption{Total amount of solid material available from 0.3 AU to 4.0 AU in a disk with a particular C/O ratio.}
    \label{tab:disk_total_masses}
    \begin{tabular}{cc}
    \hline
    \hline
         C/O Ratio & M$_{tot}$ ($M_{\oplus}$)\\
         \hline
         0.5 & 40.00\\
         0.7 & 38.96\\
         0.8 & 36.57\\
         0.9 & 35.33\\
         1.0 & 37.01\\
         1.1 & 40.10\\
         1.5 & 51.55\\
         2.0 & 62.20\\
         \hline
    \end{tabular}
    \tablecomments{We produced these values by numerically integrating our total surface density profiles for each C/O ratio. Details about this process can be found in Section \ref{sec:initial_disks}.}
\end{table}

One important assumption we made when building our disks in Section \ref{sec:initial_disks} is that all of them contained a total mass of 20 $M_{\oplus}$, even though the top panel of Figure \ref{fig:surface_density_plot} shows discrepancies between the surface density profiles for different C/O ratios. In order to show how these differences affect the total mass available in each disk, we numerically integrate our surface density profiles from 0.3 AU to 4.0 AU and place the resulting masses in Table \ref{tab:disk_total_masses}. The table shows that disks with a C/O ratio of 0.9 have the lowest amount of available mass. The table also shows that the total mass increases more rapidly from this minimum moving towards higher C/O ratios than lower ones, indicating that systems highly enriched in carbon could form higher-mass planets than silicate-dominated ones, which was mentioned in \citet{Shakespeare2025}. However, our results in Section \ref{sec:planet_compositions} show that disks with a smaller C/O ratio of 0.5 formed some of the largest planets on average, which could mitigate this effect to some degree. Still, future work should remove this assumption when performing simulations.

In addition, we applied the surface densities from one of the earliest stages of planet formation (dust condensation) to the last stage of planet formation (giant impacts). In between those stages, processes such as radial mixing and pebble accretion could significantly affect the total surface density of the disk as well as the individual surface densities of different elements and chemical species. Future studies should find ways to include the influence of other stages of planet formation when initializing late-stage integrations.

\section{Conclusions}
\label{sec:conclusions}

For this study, we performed N-body simulations of late-stage planet formation for systems with different C/O ratios with the goal of finding observable differences between silicon-rich and carbon-rich planets. C/O ratios of 0.5, 0.7, 0.8, 0.9, 1.0, 1.1, 1.5, and 2.0 were chosen; the lower four we assumed formed all silicate-dominated planets and the upper four we assumed formed all carbon-dominated planets. The simulations included a Sun-like star and 276 objects: 26 planetary embryos and 260 planetesimals which are ten times less massive than embryos.

We used two types of disks with each C/O ratio: one type that extended from 0.3 AU to 4.0 AU, which we called regular disks, and one type that extended from 0.09 AU to 0.8 AU, which we called compact disks. We determined the initial layouts of each disk for these simulations using surface density profiles of different C/O ratios obtained from a sequential dust condensation model \citep{Li2020, Shakespeare2025}. In addition, we determined the initial CMFs of objects in each disk using the surface density profiles of siderophile elements obtained from the same model. 

Ten simulations were performed for each C/O ratio and each type of disk, giving us a total of 160 planetary systems to analyze. These simulations included imperfect collisions between objects. With our results, we first studied the architectures of these planetary systems to determine if there were any significant differences between systems with carbon-rich planets and silicon-rich planets. We then determined the bulk densities of planets formed in these simulations using the interior modeling code \magrathea. We then compared these densities for different C/O ratios in order to find if this property could be used to discern between carbon-rich and silicon-rich planets in observations.

Our data showed that, across all C/O ratios, larger planets formed towards the center of a disk, a trend that was found in a previous study with a similar setup \citep{Ferich2025}. However, after analyzing the orbital properties of these systems, we found no significant differences between carbon-rich and silicon-rich planetary systems. Our results also showed that planetary CMFs tend to remain close to the initial average CMF for a particular C/O ratio, especially when it comes to larger planets. This result was similarly found in \citet{Ferich2025}. After finding the average planetary bulk densities with these CMFs using \magrathea, we determined that they are substantially different for carbon and silicon planets.

Our results indicate that bulk densities could be used to differentiate between carbon and silicon planets. However, it may not be a robust metric on its own and should be combined with other tracers --- such as the stellar C/O ratio --- to determine if a planet could be carbon-dominated. We note the setup for our simulations of late-stage planet formation was directly obtained from models of one of the earliest stages of planet formation. Future studies should determine how the stages in between these two affect the layout of material in a protoplanetary disk. 

\begin{acknowledgments}

Computational resources were provided by UNLV's National Supercomputing Center. This work is supported in part by National Science Foundation, USA grant AST-1910955. Surface density data was produced using dust condensation software from \citet{Li2020}. N-body simulations were performed with REBOUND (\url{https://rebound.hanno-rein.de}). These simulations included imperfect collisions between objects using to the fragmentation module for REBOUND (\url{https://github.com/ANNACRNN/REBOUND_fragmentation}). The final CMFs of objects from these simulations were determined using the Differentiated Body Composition Tracker (\url{https://github.com/Nofe4108/Differentiated_Body_Composition_Tracker}). \magrathea\ was used to model the interiors of silicate and carbon planets (\url{https://github.com/Huang-CL/Magrathea}). Both Numpy (\url{https://numpy.org}), Matplotlib (\url{https://matplotlib.org}) were used extensively for our analysis and plots. Finally, Google Gemini 2.5 Flash was used during the formatting for Figure 6 and Figure 7. The data shown in those figures were produced without the use of an LLM.

\end{acknowledgments}

\bibliographystyle{aasjournal}
\bibliography{references}{}

\end{document}